\documentclass[aps,pre,twocolumn,showpacs,groupedaddress,floatfix]{revtex4}

\usepackage{color,graphicx}
\usepackage{amsmath}
\newcommand{\D}[2]{\frac{\partial #1}{\partial #2}}

\begin{document}

\title{Failure of a granular step}
\author{Saloome Siavoshi and Arshad Kudrolli}
\affiliation{Department of Physics, Clark University, Worcester, MA 01610, USA}

\date{\today}

\begin{abstract}
We investigate the gravity driven rapid failure of a granular step composed of non-cohesive steel beads. The step is initially held together with electromagnets, and released when the current is switched off. We visualize the surface and the motion of the grains during the entire relaxation. The initial failure occurs at the surface and the subsequent flow is also confined to the surface as the step relaxes to its final state. The final shape of the surface is almost linear, depends on the initial angle of the step, and is not sensitive to the size of the grains. The average final slope of the pile is only slightly lower than the angle of repose of a pile formed by slowly pouring particles on to a flat surface. The evolution of the step is compared with a proposed convective-diffusion model of our system. The qualitative features of the relaxation are captured by the model after a flow-dependent dissipation parameter is introduced. \end{abstract}


\maketitle

\section{Introduction}
How a granular pile fails is important in understanding a diverse range of physical phenomena such as landslides, breaking of retaining walls and even sand castles~\cite{dade98,nedderman92,jaeger89,daerr99}. The nature of the failure depends on the applied constraints and the rate of failure. For example, in case of a retaining wall which holds granular matter,  Coulomb's method of wedges is used to analyze the failure when the wall shifts slowly and the material undergoes quasi-static failure~\cite{nedderman92}. This analysis technique assumes a planar slip surface inclined at an angle which is determined by the weight of the material, friction between the grains and the confining force exerted by the retaining wall. If the failure is catastrophic and the wall completely disintegrates, the materials accelerates rapidly and the resulting evolution can be rather different.  

Modeling rapid failure is challenging because of the existence of a broad range of contact times between particles, the transient nature of the dynamics, and a lack of a robust granular theory which encompasses both dynamic and static regimes. The situation is further complicated by a lack of quantitative data. A full scale simulation is difficult because of the wide range of grain-grain contacts that can exist during the failure. In the last decade, a continuum model has been developed by Bouchaud, Cates, Ramprakash, and Edwards (BCRE) for thin surface flows~\cite{bouchaud94}. This approach greatly simplifies the complexity of the problem by assuming a static region and a thin dynamic region flowing with constant speed, and yields a convective-diffusion equation which can be then solved with appropriate boundary conditions. The model has excited much theoretical interest~\cite{boutreux97,mahadevan99,rach02}, and has been extended to thick flows as well~\cite{boutreaux98}. 

Boutreux and de Gennes~\cite{boutreux97} have applied the BCRE equations to the evolution of a granular step. They predict a final angle of inclination of the surface which is lower than the angle of repose for a pile formed by pouring grains on a flat surface with a low rate. However, few observations of internal flow exist, and the application of the BCRE approach to this situation is based largely on conjecture.  

Recent experimental investigations have studied such a situation when a cylindrical pile is released as a retaining wall is rapidly lifted~\cite{daerr99,lajeunesse04}. These studies have focused on the dependence of the shape of the surface on the preparation of the pile, and the aspect ratio of the pile. However, the flow itself has not been well characterized which is important in developing an understanding of rapid failure. For instance, one would like to know where the initial failure occurs, i.e. does it develop at the surface or at depth. Furthermore, knowledge of the velocity fields inside the pile is important to determining the dissipation in the system.  

In this paper, we report an experimental investigation of the relaxation of a granular step using high speed imaging to address some of these open issues. Our technique allows us to measure the motion of the grains inside the pile. A key innovation is the use of magnetic fields to hold the initial step in place which results in a well controlled release with zero initial velocity. We find that the flow during the relaxation is confined mostly near the free surface. Our experiments show that the final angle of repose does not depend strongly on the initial angle of inclination, and in fact is similar to a pile obtained by pouring particles at a low rate. However, the detailed surface profile of the surface varies slightly for each case. For a pile formed by pouring at a low rate, we observe a linear surface. But for a step, the final profile shows a logarithmic deviation from linearity. We also compare the data with the model of Boutreux and de Gennes~\cite{boutreux97} and find qualitative agreement after introducing a flow-dependent dissipation parameter.  

\begin{figure}
\caption{(a-h) A sequence of velocity fields of the grains superposed on images of the step after its release at $t = 0$\,s. Velocity fields are measured by direct particle tracking. (a) $t =0.0$\,s, (b) $t = 0.04$\,s, (c) $t = 0.085$\,s, (d) $t = 0.125$\,s, and (e) $t = 0.175$\,s, (f) $t = 0.25$\,s, (g) $t = 0.45$\,s, (h) $t = 1.50$\,s. The direction and length of the vectors corresponds to the mean velocity of the particle at that location (see text). ($d = 2.8 \pm 0.2$\,mm, $h = 15$\,cm).}
\label{vfield}
\end{figure}

\section{Experimental Apparatus}
Figure~\ref{vfield}(a) shows an example of a granular step. The step is created with steel beads between two parallel plates separated by width $W = 2.5$\,cm. The pile is initially held in place with the force generated by a set of electromagnets placed behind the apparatus. The top of the pile is flat and the right side of the step has an initial angle $\theta_i = 90$\,degrees. The height of the pile $h$ is 15\,cm, in this case. The length of the step is such that the step can be considered to be semi-infinite in extent. Beads with diameters $d = 0.9 \pm 0.1$ mm, $1.7 \pm 0.2$ mm, and $2.8 \pm 0.3$ mm that are spherical to within 20\% are used in the experiments. A distributed filling procedure is adopted to uniformly build the pile. The particles are randomly packed and do not show crystalline order even near the side walls due to the irregularity in the shape and the size distribution of the particle. 

The relaxation of the step is actuated by switching off the electromagnets, and we have checked that the particles do not show any effects of residual magnetization. This technique allows us to release all the particles simultaneously. Furthermore, the particles are not susceptible to shearing that can occur while using a mechanical gate. A layer of 2.8~mm beads is glued to the bottom surface over which particles move as the pile relaxes. We image the relaxation of the pile through the transparent glass front wall using a mega-pixel resolution digital camera with a frame rate of $2,000$ per second. The entire pile is imaged with sufficient resolution to track the motion of individual particles to sub-pixel accuracy using a centroid technique~\cite{croc96}. Thus we obtain the shape of the surface and the velocity field during the entire relaxation process. 

\section{The location of failure} 
The failure of a step is shown in Fig.~\ref{vfield}. The initial pile is observed to deform rapidly initially, and then comes to rest with the free surface inclined at an angle to the horizontal. It can be noted that the surface is clearly nonlinear at early times, with the top right corner remaining recognizable as it falls through more than half the total height. After this time, the surface appears more or less linear but systematic deviations can be also noted. 

Superposed on the shape of the pile are the mean velocity fields of the grains at that instant. Here the data is plotted for clarity after measuring the grain motion over a $0.5$ ms time interval and averaging the obtained velocities over a $2.1 d$ by $2.1 d$ grid. It can be seen that the grains near the free surface flow with the greatest velocity. In fact, soon after the magnetic constraint is removed at $t = 0$ s, it can be noted that the velocity decreases to zero within a few grain diameters as one moves from the surface to the bulk [see Fig.~\ref{vfield}(b)]. Thus at early times, the flow is localized  near the free surface and the location of failure is very different from that for a quasi-static removal of the confining wall. At later time the flow gets deeper but is still confined mostly to the surface. 

To clarify the magnitudes of the relative velocities as a function of distance from the surface, we plot the mean velocity $V$ recorded at the center of the surface ($h/2$) in Fig.~\ref{vel}(a) as a function of depth $z$ normal to the surface at various time instances. At each time instance, the velocity appears to decay approximately linearly with depth. At the earliest times, barely any motion is observed for depths greater than a couple of grain diameters. As time progresses and the step relaxes, the overall speeds increase rapidly before reaching a maximum and then decrease somewhat slowly back to zero. 

\begin{figure}
\caption{(a) The velocity $V$ as a function of depth $z$ for $\theta_i = 90$ degrees. The dashed lines shows the increasing phase with $t = 0.01$ s, $t = 0.05$ s, $t = 0.085$ s, and $t = 0.125$ s and the solid lines shows the decreasing phase with $t = 0.16$ s, $t = 0.2$ s, $t = 0.29$ s, and $t = 0.44$ s. The velocity is observed to decrease smoothly to zero. (b) The maximum velocity at the surface $v_s$ as a function of time. The dashed line has a slope equal to the acceleration due to gravity. ($h = 14$ cm). (c) The corresponding depth of the flow.}
\label{vel}
\end{figure}

The corresponding speed of the beads at the surface and the depth at which the velocity decreases to zero is plotted as a function of time in Fig.~\ref{vel}(b) and Fig.~\ref{vel}(c), respectively. The data corresponds to a point approximately mid-way up the inclined surface. A line with a slope corresponding to that for gravitational acceleration is also plotted. It can be noted that for $\theta_i = 90$\,degrees, the initial acceleration of the beads is similar to gravitational acceleration but decreases before turning negative as dissipation becomes increasingly important. From these plots, we also observe that the depth of flow grows initially reaches a maximum and then decreases. We have also added data corresponding to $\theta_i = 45$\,degrees to test the effect on the initial angle on the relaxation. The overall behavior is similar except that the velocities are lower, and the peak occurs at a later time because the gravitational acceleration is lower.  

\begin{figure}
\caption{(a) A plot of the final shape of the surface as a function of the horizontal axis $x$ and the vertical axis $y$ normalized by the height of the pile. For a poured heap, we observe a linear profile, whereas the shape for the step with $\theta_i = 90$ degrees is described by the equation $ay/h + b \ln(y/h) + c = x/h$. Inset: the angle of repose of the pile as a function of the width of the container. (b) The horizontal displacement versus the vertical displacement of the center of mass of the pile for various combinations of step height and particle diameters. }
\label{h_s}
\end{figure}

\section{Evolution of the free surface} 
We first discuss the final shape of the pile before examining the change in its shape during the relaxation. The final surface profiles are plotted in Fig.~\ref{h_s}(a) for three different initial conditions. The profiles obtained for $\theta_i = 90$\,degrees, and $\theta_i = 45$\,degrees are plotted along with the shape for a pile formed by pouring the grains with a slow rate of about 20 grains per second inside the same container in which the relaxation experiments are conducted. The grains are poured from a point only slightly above the surface pile to prevent the particles from gaining substantial kinetic energy before they impinge on the pile surface. Because the surface flow can be intermittent when the grains are poured slowly, we examine the surface profile for the pile right after an avalanche event has occured to be consistent. The purpose of these second kind of experiments is to provide a baseline to compare the effects of the rapid motion, and to estimate the effects of the side walls on the observed shapes. As can be noted from Fig.~\ref{h_s}(a), the steps relax only slightly further than the piles formed by slow pouring. The relaxation for $\theta = 90$\,degrees is systematically greater than for $\theta = 45$\,degrees.  

We find that the pile surface formed by slow pouring is described by a straight line. By contrast, the final shape after a rapid relaxation from an initial step shape is best described by an equation containing a linear and a logarithmic term: $x/h = ay/h + b \ln{(y/h)} + c$, where $y$ is the height of the surface at a horizontal position $x$, and $a$, $b$, and $c$ are constants. For the case shown in Fig.~\ref{h_s}(a), $a = -1.865$, $b = -0.050$, and $c = 1.877$, for $\theta_i = 90$\,degrees, and $a = -1.764$, $b = -0.014$, and $c = 1.779$, for $\theta_i = 45$\,degrees. We also varied the height $h$ of the pile between 8 cm and 20 cm and found that $a$, $b$, and $c$ remains unchanged with standard deviations within 0.075, 0.021, 0.083, respectively.  

Our results for the shapes of the piles formed by slow pouring are inconsistent with the conditions reported in previous reports~\cite{alonso96,jin04}. In these previous reports, logarithmic deviations from linearity were reported for piles formed by pouring and compared with models where dynamics was neglected. However, we observe logarithmic deviations either after a rapid step relaxation or  when grains are dropped on to the pile from a height so that they have substantial kinetic energy when they impact the pile. Thus we observe logarithmic deviations only when particles have substantial kinetic energy during the relaxation. 

In order to examine the effect of the side walls, we measured the surface of piles formed by pouring beads inside containers with various combinations of widths $W$ and bead sizes. The surface was observed to be linear in all cases. The corresponding angle of repose as a function of the width of the system normalized by the particle diameters is shown in the inset to Fig.~\ref{h_s}(a). The dashed line is an exponential fit to the data. The asymptotic value for the angle of repose is $25.5$\,degrees, and thus side walls do have a small effect on the measured surface profiles, but do not alter the qualitative features.  

As noted earlier, the final shape of the pile does not appear to depend significantly on step height. To illustrate the scaling of the  relaxation, we plot the horizontal $\Delta x$ versus vertical $\Delta y$ displacement of the center of mass of the pile with data obtained for three different particle sizes [see Fig.~\ref{h_s}(b)]. We plot the displacement of the center of mass rather than the length of the relaxation because it is more suseptible to fluctuations in the tail of the relaxation. A linear dependence is observed. Thus, it appears that the surface profile is not very sensitive to the grain size, which gives support to a continuum approach to describing the relaxation.  

\section{Comparison with a convection-diffusion model}
As first discussed in the introduction, a phenomenological continuum description of this situation has been developed by Boutreux and de Gennes~\cite{boutreux97} based on the BCRE model. They assumed that the flow during relaxation is confined to the surface consistent with our observations. We next briefly reproduce their argument for the rate of change of the surface as the step relaxes before comparing the predictions with our data. They first assumed that the surface is approximately linear, and given by $y = x \tan{\theta}$. Then it was argued that the rate of change of the total height of the pile is given by
\begin{equation}
\frac {dy}{dt} \approx v \D{R}{x},  
\label{dh}
\end{equation}
where, $R$ is the height of the liquid-like mobile phase, and $v$ is the mean velocity of particles in the mobile phase. Next they assumed that $R$ has a maximum $R_0$ at the center and decreases smoothly to zero at the front and back edge of the step and therefore
\begin{equation}
\D{R}{x} = \alpha \frac{R_0 \tan{\theta}}{h},  
\label{dR}
\end{equation}
where, $\alpha$ is assumed to be a non-dimensional number with an absolute value of order 1 near the extremum and less than 1 elsewhere. Next they assume that the rate of change of the mobile phase at the maximum mobile depth $R_0$ at the step-center is given by
\begin{equation}
\frac{dR_0}{dt} = \gamma (\tan{\theta} - \tan{\theta_d}) R_0,
\label{dR0}
\end{equation}
where, $\gamma$ is a frequency which is related to the flow velocity $v$ and the diameter of the particle. They assumed $\theta_d$ to be the angle of repose of the particles $\theta_r$. This equation comes from the BCRE model~\cite{bouchaud94,mahadevan99} and it is further assumed that convection has negligible effect on the rate of change of the rolling phase at its maximum. 

Next, they substituted Eq.~(\ref{dR}) in Eq.~(\ref{dh}) to get $R_0$ which in turn is replaced in Eq.(\ref{dR0}) to give a second order differential equation for the distance traveled $s$ after rewriting the variable $y$ in terms of $s$. This equation upon integration yields the rate of change of the runoff distance $s$ normalized by the initial height of the step $\cot{\theta} = s/h$~\cite{boutreux97}: 
\begin{equation}
\frac {1}{\gamma} \frac{d\cot{\theta}}{dt} = \ln(\cot{\theta}) - \tan{\theta_d} \cot{\theta} + k,
\label{crit}
\end{equation}
where, $k$ is a constant of integration. Equation~(\ref{crit}) predicts a maximum for $\frac{d\cot{\theta}}{dt}$ at $\theta=\theta_d$ which can be tested against our experimental results. 

\begin{figure}
\caption{(a) The evolution of the average inclination angle of the granular free surface. (b) The normalized runoff distance of the step $\cot{\theta} = s/h$ as a function of time. The dashed lines are the numerical integrations of Eq.~(4) (see text). (c) The rate of change of the normalized runoff distance $\cot{\theta} = s/h$. The dashed lines are fits to Eq.~(4). The maximum occurs at higher angles for the higher initial angles, and seems to be independent of the initial height of the step. (d) Velocity dependence of the effective dissipation for the particles at the surface.}
\label{ds}
\end{figure}

Figure~\ref{ds}(a) shows the plot of the average surface angle of the pile corresponding to a step with three different initial angles $\theta_i$ = 45 degrees, 60 degrees, and 90 degrees. Because the surface is not quite linear as discussed earlier, we average the angle over the entire free surface but exclude 3 particle diameters from the top and bottom of the step. The data shows a rapid decreases at early times with a crossover to a slower decrease to the final angle. In Fig.~\ref{ds}(b), we plot the corresponding normalized runoff distances $\cot{\theta}$ as a function of time. Then to compare the data to the model, the rate of change of $\cot{\theta}$ for the same data is plotted in Fig.~\ref{ds}(c). The peak occurs well before $\theta_r$ is reached in all cases and depends on the initial angle. Thus if $\theta_d=\theta_r$ as assumed in the discussion of the model above, then even qualitative agreement is not seen. 

Next we allow $\theta_d$ to vary and find that Eq.~(\ref{crit}) can indeed qualitatively describe the data. In Fig.~\ref{ds}(c), the dashed lines are the fits to Eq.~(\ref{dR0}). The corresponding fitting parameters are listed in Table ~\ref{coef}. This possibility was also discussed by Boutreaux, {\em et al.}~\cite{boutreaux98} while extending the BCRE equations to thick flows. They anticipated that $\theta_r$ in Eq.~\ref{dR0} may have to be replaced by a higher neutral angle. Our observation of an higher $\theta_d$ is also consistent with experiments on heap flows where an increase in neutral angle is observed with flow rate~\cite{khakhar01}. 

\begin{table}
\begin{center}
\begin{tabular}{ccccc}
\hline
$\theta_i$ (deg.) & h (cm) &  $\gamma$ ($s^{-1}$) & $\theta_d$ (deg.) & $k$ \\
\hline
\hline
90 & 17.5 & 1.89 & 68.4 & 3.947 \\
60 & 17.5 & 15.26 & 43.8 & 1.156 \\
45 & 12.5 & 50.48 & 36.2 & 0.735 \\
90 & 14.0 & 3.92 & 67.6 & 3.820 \\
60 & 13.0 & 19.72 & 43.9 & 1.154\\
\hline
\end{tabular}
\caption{\label{coef} Parameters $\gamma$, $\theta_d$, and $k$ obtained by fitting Eq.~(\ref{crit}) to the data. The first three rows corresponds to the data plotted in Fig.~\ref{ds}(c).}
\end{center}
\end{table}

We note from Table~\ref{coef} that $\theta_d$ increases systematically with $\theta_i$ which results in higher flow speeds as can be noted from Fig.~\ref{vel}(b). Now we can also use the fitting constants to numerically integrate Eq.~(\ref{crit}) and obtain the normalized runoff distance as a function of time. The corresponding curves (dashed lines) are shown in Fig.~\ref{ds}(b) and qualitatively captures the evolution of the normalized runoff as well. Repeating the experiment with different heights for the same $\theta_i$ gives similar $\theta_d$. Thus we further confirm the scaling behavior of the system. 

The fact that $\theta_d$ is greater than $\theta_r$ and increases with $\theta_i$ seems to be self-consistent with the idea that dissipation in granular flows occurs not only because of sliding friction but also because of inelastic collisions.   Figure~\ref{ds}(d) shows the velocity dependence of the normalized effective dissipation force at the surface layer. This has been obtained by balancing the average forces on a particle due to gravitation acceleration along the inclined surface and the dissipative force normalized by the weight of the material $f_{dis}$ at the surface of the pile and can be simply calculated from $f_{dis} = \sin\theta - a/g$. One notes that the dissipative force is an increasing function of velocity and has a maximum at the maximum velocity before finally dropping down to the static value as the material comes to rest. The observed hysteresis arises possibly because of differences in the contribution of sliding friction during the deceleration and acceleration phase due to differences in inclination angles. However, without a detailed dissipation model for transient flows it is difficult to interpret the data further. 

\section{Conclusions}
In summary, by visualizing the failure of a granular step, we have found that it relaxes to angles only slightly less than the angle of repose of a pile formed by slow pouring. Furthermore, the flow is observed to be confined to the free surface with the velocity decaying smoothly with depth. The step is observed to relax rapidly at first with the maximum rate occuring at an angle greater than the angle of repose. Thus the resistance to motion offered by inelastic collisions in addition to sliding friction must be important in determining the evolution of the pile. From our experimental observations, it appears that a convective-diffusion model can describe the qualitative features of a rapid step relaxation by introducing a velocity dependent dissipation parameter. It is interesting that this is the case considering  that some of the assumptions in the model such as constant flow speed are not met in the experiments. Nonetheless, the fact that such an approach can capture the dynamic evolution is interesting considering the difficulty in modelling dense granular flows. Further work is required to develop and understand the physical implication of the model now that experimental data is avaliable.  

Note added - During the review of our manuscript we became aware of two papers which discuss granular column collapse in two and three dimensions~\cite{lube04,balmforth} in addition to Ref 10. These studies examine the scaling of the runoff distance with column width and may be considered complementary to our study focusing on semi-infinite exten columns.

\begin{acknowledgments}
We thank L. Mahadevan, A. Samadani, D. Blair, and A. Orpe for many insightful suggestions during the development of this project. This work was supported by the National Science Foundation under grant \# CAREER-9983659 and CTS-0334587, and by the GLUE program of the Department of Energy.
\end{acknowledgments}

\end{document}